\documentclass[a4paper,10pt,twoside]{cpc-hepnp}
\usepackage{CJK,upgreek,fancyhdr}
\usepackage{multicol}
\usepackage{graphicx}
\usepackage{booktabs}
\usepackage{amssymb,bm,mathrsfs,bbm,amscd}
\usepackage[tbtags]{amsmath}
\usepackage{lastpage}

\begin{document}
\begin{CJK*}{GB}{gbsn}

\fancyhead[c]{\small Chinese Physics C~~~Vol. xx, No. x (201x) xxxxxx}
\fancyfoot[C]{\small 010201-\thepage}

\footnotetext[0]{Received 18 March 2020}

\title{Description of alpha-clustering of  $^8$Be nucleus states in high-precision theoretical approach}

\author{%
      D. M. Rodkin $^{1;1)}$\email{rodkindm92@gmail.com}%
\quad Yu. M. Tchuvil'sky $^{1,2;2)}$\email{tchuvl@nucl-th.sinp.msu.ru}%
}
\maketitle

\address{%
$^1$ Dukhov Research Institute for Automatics, 127055, Moscow, Russia \\
$^2$ Skobeltsyn Institute of Nuclear Physics, Lomonosov Moscow State University, 119991 Moscow, Russia\\
}

\begin{abstract}
Scrupulous theoretical study of $^8$Be nucleus states, both clustered and non-clustered, is performed  over a wide range of the excitation energies. The quantities which characterize the degree of the alpha-clustering of these states: spectroscopic factors, cluster form factors as well as the alpha-decay widths are computed in the framework of an accurate ab initio approach developed. Other basic properties of $^8$Be spectrum: the binding and excitation energies, mean values of the isospin are calculated simultaneously. In the majority of instances the results of the computations turn out to be in a good agreement with the spectroscopic data. A number of predictions are made and corresponding verification experiment is proposed. Prospects of the developed approach for nuclear spectroscopy are demonstrated.
\end{abstract}


\footnotetext[0]{\hspace*{-3mm}\raisebox{0.3ex}{$\scriptstyle\copyright$}2013
Chinese Physical Society and the Institute of High Energy Physics
of the Chinese Academy of Sciences and the Institute
of Modern Physics of the Chinese Academy of Sciences and IOP Publishing Ltd}

\begin{multicols}{2}

\section{Introduction}

One of the fundamental properties of nuclei are the clustering phenomena i.e. the effects which present themselves in a certain degree of separation of a nucleus into two or more multi-nucleon substructures. These phenomena are displayed  in internal properties of nuclei and intimately connected with the properties of nuclear reactions with composite particles in entrance and/or exit channels as well as with characteristics of nuclear resonances decay.  

The elaboration of a microscopic, i. e. starting from a certain NN-potential and considering a nucleus or a two-fragment collision channel as an A-nucleon or an (A$_1$ + A$_2$)-nucleon system theoretical  method resulted in the emergence of the Resonating Group Model (RGM)  \cite{rgm1, rgm2}. In succeeding years a variety of additional theoretical techniques have been developed to study nuclear clustering such as the Generator Coordinate Method (GCM) \cite{hor,desc1}, Microscopic Cluster Model \cite{desc2,desc3}, THSR-approach \cite{thsr}, Antisymmetrized Molecular Dynamics (AMD) \cite{keh},  Algebraic Version of the RGM \cite{av1, av2} and Cluster-Nucleon Configuration Interaction Model \cite{avil,vt2,vt3,vt4}. In this models various aspects of clustering phenomena are studied. These methods are adapted for studying cluster effects in light nuclei. It should be noted that the most of these methods are based on effective NN-potentials and used for calculations of specific highly clustered states. 

Modern high-precision methods for describing both light nuclei properties and characteristics of reactions 
induced by light nuclei collisions are advancing nowadays. Such approaches, being called ab initio, are based on
new possibilities provided by modern high-performance supercomputers. An important role
among  the methods describing light nuclei structure belong to various ab initio methods,
such as different versions of No-Core Shell Model (NCSM) \cite{ncsm1, ncsm2, ncsm3, dyt1, dr},
Gamov Shell Model (GSM) \cite{gamov1, gamov2,gamov3}, Green functions Monte Carlo method \cite{gfmc1,
gfmc2, gfmc3}, the Coupled Cluster Method \cite {ccm1} and Lattice Effective Field
Theory for Multi-nucleon Systems \cite{left1, left2, left3}. These methods are all based
on realistic NN-, NNN-, etc. potentials. These potentials could be derived from Chiral
Effective Field Theory \cite{dj16, machleidt1, machleidt2} or from nucleon
scattering data by the use of $J$-matrix inverse scattering method \cite{jisp1}. 

In the current work a new metod of such a type is developed.  It is based on one of the versions of NCSM and a special projecting of its wave functions (WFs) technics. The Daejeon16 NN-potential \cite{dj16} which is built using the N3LO limitation of Chiral Effective Field Theory \cite{ceft1} softened by similarity renormalization group (SRG) transformation \cite{srg1} and the JISP16 NN-potential, which is based on inverse scattering method \cite{jisp1} are used. These potentials were tested in large-scale calculations of different properties of nuclei with mass $A \le 16.$ and showed their reliability. Naturally the more recent Daejeon16 potential gives results of higher quality in this mass region due to both more accurate fit of its parameters and a better convergence of the variational  procedure. We carry out the computations of the eigenvalues and the eigenvectors of discussed Hamiltonians using NCSM. This approach is one of the most advanced and reliable among various ab initio methods. This model is based on solving A-nucleon Schr$\ddot o$dinger equation using realistic NN-potentials on the complete basis of totally antisymmetric A-nucleon WFs up to the maximal total number of oscillator quanta N$^{max}_{tot}$. The size of this basis, for example, in widely met M-scheme reaches sometimes the value of about 10$^{9}$ -- 10$^{10}$ in the case that a modern supercomputer is employed. This method was used for calculations of the binding and excitation energies characterizing ground and excited states of nuclei and unstable resonances, as well as the nuclear sizes and their electromagnetic observables in a lot of works.

NCSM model and methods similar to it  are, however, not adapted to describe clustering effects, nuclear reactions and asymptotic properties of nuclei resonances directly. For this purposes different methods were proposed. For systems with $A \le 5$ ab initio description of continuum spectrum states could be based on Faddeev and Faddeev-Yakubovsky equations \cite{fy1, fy2}. 
Ab initio approaches focused on the discussed problem are also presented in the literature.
Among them the methods which combine NCSM and RGM namely No-Core Shell
Model / Resonating Group Model (NCSM/RGM) \cite{ncsmrg1} and No-Core Shell Model with
Continuum (NCSMC) \cite{ncsmc1, ncsmc2, ncsmc3, ncsmc4} seem to be the most versatile. Examples of 
a good description of the asymptotic  characteristics of decay channels of the spectra of $^7$Be
 and $^7$Li are presented in  \cite{ncsmc4}.  
As the NCSMC the Fermionic Molecular Dynamics (FMD) \cite{neff1,neff2,neff3} offers in fact an ab initio
approach focussed on the unified description of both bound states and continuum ones. 
Another approach -- Cluster Channel Orthogonal Functions Method (CCOFM) -- is also based on the employment of a basis 
combing NCSM and orthogonalized cluster-channel WFs \cite{our1, our2}. 

Alpha clustering is undoubtedly the most common and important  cluster phenomenon. Alpha decay of nuclear resonances is, together with nucleon emission, the most popular decay process in the experimental nuclear spectroscopy. Nevertheless ab initio theoretical calculations of the process are presented in the literature by a fairly small number of works. In Ref. \cite{our2} the alpha-cluster properties of the rotational band of $^8$Be nucleus was studied. The alpha-particle spectroscopic factors (SFs) but not alpha-decay widths were presented. The alpha decay of the states of the same band was the subject of the Refs.  \cite{kv1,kv2,kv3}.  Another version of RGM based on the realistic NN-potentials including the JISP16 (in Ref. \cite{kv3}) was put into use for calculations involving a bases limited by the maximal total number of oscillator quanta N$^{max}_{tot}$=12  \cite{kv1}  10 or \cite{kv2,kv3} respectively. NCSM calculations were also performed but their results were used for the computations of the decay widths indirectly -- as a control method for the results of RGM calculations of the excitation energies and the SFs.   

In the current paper the binding and the excitation energies as well as the alpha-decay properties of all positive parity states, both clustered and non-clustered, located in a wide range of  $^8$Be nucleus excitation energies are studied in a unified scheme. The scheme is based on the computations of the Hamiltonian eigenvalues and eigenvectors in the framework of conventional NCSM involving the complete  basis with the cut off parameter N$_{max}^{tot}$ = 14 and projection of the resulting  eigenvectors onto the WFs of the cluster channels obtained in the framework of CCOFM. The values obtained in the framework of the projecting procedure namely the alpha-particle form factors serve to calculate the decay widths of the states under study. The binding and excitation energies of the states under study as well as the statistical weights of the components with the isospin $T$=1, which are also basic characteristics of the nucleus spectrum, constitute an integral part of of the complex of studied objects together with the alpha-decay widths.

\section{Formalism of calculating cluster spectroscopic factors,  cluster form factors and cluster decay widths}

Let us demonstrate how translationally invariant A-nucleon WFs of arbitrary two-fragment decay channel with
separation A = A$_1$ + A$_2$ are built in CCOFM. The useful  feature of the procedure is that 
each function of this basis can be represented as a superposition of
Slater determinants (SDs). To do that the technique of so-called cluster coefficients (CCs) is exploited. 

The oscillator-basis terms of the cluster channel $c_\kappa$ are expressed in the following form:
\begin{equation}
\Psi^{c_\kappa} _{{A\,},nl}  = \hat A\{\Psi^{\{k_1\}} _{A\,_1 } \Psi^{\{k_2\}} _{A\,_2 }
\varphi _{nl} (\vec \rho )\} _{J_cJM_JT} , \label{eq0}
\end{equation}
 where  $\hat A$ is the antisymmetrizer, $\Psi^{\{k_i\}}
_{A\,_i}$ is a translationally invariant internal WF of the fragment labelled by a set
of quantum numbers $\{k_i\}$; $\varphi _{nlm} (\vec \rho )$ is the WF of the relative
motion. The channel WF is labelled by the set of quantum numbers $c_\kappa$ which
includes $\{k_1\},\{k_2\},n,l,J_c,J,M_J,T$, where $J$ is the total momentum and $J_c$ is
the channel spin.  As it was mentioned above the function has to be represented
as a linear combination of the SDs containing the one-nucleon WFs
of the oscillator basis. For these purposes two-fragment WF in another representation
\begin{equation}
\Psi^{\{k_1,k_2\}} _{{A\,},nlm}  = \hat A\{\Psi^{\{k_1\}} _{A\,_1} \Psi^{\{k_2\}}
_{A\,_2} \varphi _{nlm} (\vec \rho )\} _{J_c,M_{J_c},M_JT}  \label{eq1}
\end{equation}
is multiplied by the function of the center of mass (CM) zero vibrations $\Phi _{000}
(\vec R)$. Then the transformation of WFs caused by changing from $\vec R,\vec \rho$ to
$\vec R_1 ,\vec R_2$ coordinates -- Talmi-Moshinsky-Smirnov transformation
 is performed \cite{talmi-moshinsky} and WF (\ref{eq1}) takes the form

\begin{equation}
\begin{array}{rcl}
\Phi _{000} (\vec R)\Psi^{\{k_1,k_2\}} _{{A\,},nlm} = \sum\limits_{N_i ,L_i ,M_i}
{\left\langle {{\begin{array}{*{20}c}
   {000}  \\
   {nlm}  \\
\end{array}  }}
 \mathrel{\left | {\vphantom {{\begin{array}{*{20}c}
   {000}  \\
   {nlm}  \\
\end{array}  } {\begin{array}{*{20}c}
   {N_1 ,L_1 ,M_1 }  \\
   {N_2 ,L_2 ,M_2 }  \\
\end{array}}}}
 \right. \kern-\nulldelimiterspace}
 {{\begin{array}{*{20}c}
   {N_1 ,L_1 ,M_1 }  \\
   {N_2 ,L_2 ,M_2 }  \\
\end{array}}} \right\rangle } \\[\bigskipamount]
\hat A\{ \Phi _{N_1 ,L_1 ,M_1 }^{A_1 } (\vec R_1 ) \Psi^{\{k_1\}}_{A\,_1 } \Phi _{N_2
,L_2 ,M_2 }^{A_2 } (\vec R_2 )\Psi^{\{k_2\}} _{A\,_2 } \} _{J_c,M_{J_c},M_JT} . \label{eq2}
\end{array}
\end{equation}

The main procedure of the method is to transform internal WFs corresponding to each
fragment multiplied by none-zero center of mass vibrations into a superposition of SDs

\begin{equation}
\Phi _{N_i ,L_i ,M_i }^{A_i } (\vec R_i )\Psi^{\{k_i\}}_{A\,_i }  = \sum\limits_k
{X_{N_i ,L_i ,M_i }^{A_i (k)} \Psi _{A\,_i (k)}^{SD} }. \label{eq3}
\end{equation}
Quantity $X_{N_i ,L_i ,M_i }^{A_i (k)}$ is called a cluster coefficient (CC). Technique of
these objects is presented in detail in \cite{nem}. There is a large number of methods
elaborated for the calculations of CCs. The most general scheme is based on the method
of the second quantization of the oscillator quanta. In this scheme the WF of the CM
motion is presented as

\begin{equation}
\Phi _{N_i ,L_i ,M_i }^{A_i } (\vec R_i ) = N_{N_i ,L_i } (\hat{ \vec \mu^{\dag}})^{N_i
- L_i } Y_{N_i ,L_i } (\hat{ \vec \mu^{\dag}})\Phi_{000}^{A_i} (\vec R_i ), \label{eq4}
\end{equation}
where  $\hat{ \vec \mu^{\dag}}$ is the creation operator of the oscillator quantum, and
$N_{N_i ,L_i }$ is the norm constant. The creation operator of oscillator quanta of the center of mass vibrations is represented as follows:

\begin{equation}
\hat{\vec \mu^{\dag}} = \frac{1}{\sqrt{A}} \sum_{i=1}^{A} \vec a_{i}^{+}. \label{eq4}
\end{equation}
Thus the CC turns out to be reduced to a matrix
element of the tensor operator expressed in terms of $ \hat{ \vec \mu^{\dag}}$:

\begin{equation}
\begin{array}{rcl}
 < \Psi _{A_i (k)}^{SD} |\Phi _{N_i ,L_i ,M_i }^{A_i } (\vec R_i ) \Psi^{\{k_i\}}_{A_i }  >  =
 N_{N_i ,L_i }
  \left\langle {\Psi _{A\,_i (k)}^{SD} }
 \right|  \\
 (\hat \mu ^\dag  )^{N_i  - L_i }
 Y_{N_i ,L_i } (\hat{ \vec \mu^{\dag}})
 \left| {\Phi _{000}^{A_i } (\vec R_i )\Psi^{\{k_i\}}_{A\,_i } } \right\rangle
\end{array}
\label{eq5}
\end{equation}

A conventional relationship between the a translationally invariant and  an ordinary
shell-model WFs
\begin{equation}
\Psi _{A\,_i }^{shell} = \Psi^{\{k_i\}}_{A\,_i } * \Phi _{000}^{A_i } (\vec R_i )
\label{eq6}
\end{equation}
is used as a definition of the former one. The NCSM WFs of the fragments $\Psi _{A\,_i
}^{shell}$ are involved in the calculations.

For the calculations of $\Phi _{N_i ,L_i ,M_i }^{A_i } (\vec R_i )\Psi^{\{k_i\}}_{A\,_i }$ WFs with N+1 oscillator quanta along the center of mass coordinate the total set of WFs $\Phi _{N_i ,L_i ,M_i }^{A_i } (\vec R_i )\Psi^{\{k_i\}}_{A\,_i }$ with N oscillator quanta  is used in the following set of equations:

\end{multicols}

\begin{equation}
\begin{split}
\hat{\mu^{\dag}_q}| \Phi^{A_i}_{N_iL_iM_i}(\vec R_i) \Psi^{\{k_i\}}_{A\,_i } \rangle = \frac{1}{\sqrt{A}} \sum \limits_{i=1}^{A} a^{\dag}_{iq} \sum \limits_{k} X_{N_i ,L_i ,M_i }^{A_i (k)} \Psi _{A\,_i (k)}^{SD} = \\
\frac{1}{\sqrt{2L_i+3}} C^{L_iM_i1q}_{(L_i+1)(M_i+q)} < N_i L_i+1\left\Vert \mu^{\dag} \right\Vert N_i L_i > 
 | \Phi^{A_i}_{N_i(L_i+1)(M_i+q)}(\vec R_i) \Psi^{\{k_i\}}_{A\,_i } \rangle + 
\\\frac{1}{\sqrt{2L_i-1}} C^{L_iM_i1q}_{(L_i-1)(M_i+q)} < N_i+1 L_i-1\left\Vert \mu^{\dag} \right\Vert N_i L_i > 
 | \Phi^{A_i}_{(N_i+1)(L_i-1)(M_i+q)}(\vec R_i) \Psi^{\{k_i\}}_{A\,_i } \rangle
\end{split}
\label{eq6}
\end{equation}

\begin{multicols}{2}

Using this set of equations and  the set of Talmi-Moshinsky-Smirnov coefficients  one can construct  WFs $\Phi _{000} (\vec R)\Psi^{c_\kappa} _{{A\,},nlm}$ (\ref{eq2}). The last step is to construct $\Psi^{c_\kappa} _{{A\,},n}$ basis WFs for each channel $c_\kappa$ (\ref{eq1}) from a basis of $\Psi^{\{k_1,k_2\}} _{{A\,},nlm}$ by the use of Clebsh-Gordan coefficients. 

It should be noted that WFs of terms (\ref{eq0}) of one and the same
channel $c_\kappa$ characterized by the pair of internal functions
$\Psi^{\{k_1\}}_{A_1}$, $\Psi^{\{k_2\}}_{A_2}$  and extra quantum numbers
$l,J_c,J,M_J,T$ are non-orthogonal. Creation of orthonormalized basis functions of
channel $c_\kappa$ is performed by the diagonalization of the exchange kernel

$ ||N_{nn'} || =  < \Psi^{c_\kappa} _{{A\,},n'}|\Psi^{c_\kappa} _{{A\,},n} > = $
\begin{equation}
< \Psi^{\{k_1\}} _{A_1} \,\Psi^{\{k_2\}} _{A_2} \,\varphi _{nl} (\rho )
|\hat A^2 |\Psi^{\{k_1\}} _{A_1} \,\Psi^{\{k_2\}} _{A_2} \,\varphi _{n'l} (\rho ) > . \label{eq7}
\end{equation}

The eigenvalues and eigenvectors of this exchange kernel are given by the expressions:

\begin{equation}
\varepsilon _{\kappa,k}  =  < \hat A\{ \Psi^{\{k_1\}} _{A_1} \,\Psi^{\{k_2\}} _{A_2} \,f_l^k (\rho
) \} |\hat 1|\hat A\{ \Psi^{\{k_1\}} _{A_1} \,\Psi^{\{k_2\}}_{A_2} \} \,f_l^k (\rho
)  > ; \label{eq8}
\end{equation}

\begin{equation}
f_l^k (\rho ) = \sum\limits_n {B_{nl}^k \varphi _{nl} (\rho )}. \label{eq9}
\end{equation}

As a result, the WF of the orthonormalized basis channel basis $c_\kappa$ is
represented in the form

\begin{equation}
\Psi^{SD,c_\kappa} _{{A\,},kl}=\varepsilon^{-1/2} _{\kappa,k}|\hat A\{ \Psi^{\{k_1\}}
_{A_1} \,\Psi^{\{k_2\}}_{A_2} \,f_l^k (\rho ) \}  >, \label{eq10}
\end{equation}

The cluster form factor (CFF)  is a projection of the function of an initial A-nucleon state $\Psi_{A}$ 
 onto the WF of a particular channel $c_\kappa$. It  describes the relative motion of subsystems and  has the form

$$\Phi^{c_\kappa}_A(\rho)=\sum\limits_k\varepsilon^{-1/2} _{\kappa,k}<\Psi _{A}|\hat A\{ \Psi^{\{k_1\}} _{A_1}
\,\Psi^{\{k_2\}}_{A_2} \,f_l^k (\rho ) \}>f_l^k (\rho )$$
\begin{equation}
=\sum\limits_k\varepsilon^{-1/2} _{\kappa,k}\sum\limits_{n, n'}B_{nl}^k B_{n'l}^{k}
C_{AA_1A_2}^{n'l}\varphi _{nl} (\rho ) \label{eq11}
\end{equation}
where the mathematical object 

$C_{AA_1A_2}^{nl} = < \hat A\{\Psi^{\{k_1\}} _{A_1} \Psi^{\{k_2\}} _{A_2} \varphi _{nl}
(\rho ) \} | \Psi _{A} >=$
\begin{equation}
<\Psi^{SD,c_\kappa} _{A, nl}|\Phi _{000} (R)| \Psi _{A} >=<\Psi^{SD,c_\kappa} _{A,
nl}|\Psi^{SM} _{A} >. \label{eq12}
\end{equation}
is called the spectroscopic amplitude. Very diverse methods of its
calculation depending on the masses of the initial nuclei and fragments are described
in \cite{nem,sa1, sa2, sa3}. The amplitude of the CFF is determined as

\begin{equation}
K_{AA_1A_2}^{nl}=\sum\limits_{k, n'}\varepsilon^{-1/2} _{\kappa,k}B_{nl}^k B_{n'l}^{k}
C_{AA_1A_2}^{n'l} .\label{eq13}
\end{equation}
The SF of this channel $c_\kappa$ has the form

$$S_l^{c_\kappa}  = \int {|\Phi^{c_\kappa}_A(\rho)|^2 } \rho ^2 d\rho =$$
\begin{equation}
\sum\limits_k {\varepsilon _k^{ - 1} \sum\limits_{nn'} {C_{AA_1A_2}^{nl} }
C_{AA_1A_2}^{n'l} B_{nl}^k } \;B_{n'l}^k . \label{eq14}
\end{equation}

The definitions of the CFF (\ref{eq11}) and SF (\ref{eq14}) are completely
equivalent to those proposed in \cite{newsf1} (the so-called "new" spectroscopic
factor and CFF). In contrast to the traditional definition, the new
SF characterizes the total contribution of orthonormalized cluster
components to the WF which is a solution of the Schr$\ddot o$dinger equation for A nucleons. Reasons for the
necessity of its use to describe decays and reactions can be found in \cite{newsf2,
newsf3}. In \cite{vt2, vt3,vt4}, it was demonstrated that the correct definition eliminates
a sharp contradiction between theoretical calculations of the cross sections for
reactions of knock-out and transfer of $\alpha$ clusters and experimental data.

Compared to the SF the CFF is a more informative characteristic because it
allows its matching with the asymptotic WF of the relative motion in the range of validity of shell-model solution
and thus determines the amplitude of the WF in the asymptotic region. 
 
It is convenient to use the CFF for computing the widths of resonances and the
 asymptotic normalization coefficients of bound states, which in turn are used to
calculate the cross sections of resonant and peripheral reactions respectively. The CFF in its new
definition allows matching with the asymptotic WF at relatively small
distances, where the nuclear interaction is negligibly weak, but exchange effects
generated by the antisymmetry of the total channel WF are still not negligibly small.
This property is very important for dealing with NCSM WFs.

In the proposed approach, a direct matching procedure is applied to calculate  the widths of narrow resonances. For such resonances or, more precisely, for those of them whose small width is due to a
low penetrability of the potential barrier, we used a very compact procedure proposed in
\cite{matchingpoint}. This procedure is applicable because  for such resonances there is  sufficiently
wide range of distances in which the nuclear attraction is already switched off and at the same time 
the potential barrier is high enough.
At any inner point $\rho_{in}$ of this area, the relationship between the regular and irregular solutions  $F_l (\eta,k\rho_{in} )$ and $G_l (\eta,k\rho_{in} )$ of the two-body Schr$\ddot o$dinger equation in the WKB approximation has the form

\begin{equation}
F_l (\eta,k\rho_{in} )/ G_l (\eta,k\rho_{in} ) = P_l (\rho_{in} ) \ll 1\label{eq15}
\end{equation}
where $P_l (\rho_{in} )$ is the penetrability of the part of the potential barrier which
is located between the point $\rho_{in}$ and the outer turning point. The smallness of
this penetrability is the condition of applicability of the approximation, where the
contribution of the regular solution can be neglected.
To determine the position of the matching point of the CFF and irregular WF in this
range, we use the condition of equality of the logarithmic derivatives

\begin{equation}
\frac{d\Phi _A^{c_\kappa  } (\rho)/d\rho}{\Phi _A^{c_\kappa  } (\rho )}= \frac{dG_l
(\eta,k\rho)/d\rho}{{G_l (\eta,k\rho )}}, \label{eq16}
\end{equation}
which determines the matching point $\rho_{m}$; therefore, the decay width is given
by the expression

\begin{equation}
\Gamma  = \frac{{\hbar ^2 }}{{\mu k}}\left [ \frac{{\Phi _A^{c_\kappa  } (\rho
_{m})}}{{G_l (\eta,k\rho _{m})}}\right ]^2. \label{eq17}
\end{equation}

To make the list of the states of a certain nucleus accessible for studies in the discussed scheme wider large-width
resonances are considered in the following way. In the case that a resonance is wide and so the penetrability $P_l
(\rho_{in} )$ (\ref{eq15}) is not small the width of this resonance is calculated using the
 simple version of the conventional R-matrix theory in which the decay width takes the form:

\begin{equation}
\Gamma  = \frac{\hbar^2}{\mu k} (F^2_l (\eta,k\rho_{m} ) + G^2_l (\eta,k \rho_{m} ))^{-1}
(\Phi _A^{c_\kappa} (\rho_{m} ))^2. \label{eq20}
\end{equation}
Naturally the use of this version leads to reduction in accuracy of calculation results, but it should be noted
that the accuracy of the the data, concerning large-width resonances, extracted from various experiments is also 
very limited. That is why the simplified version of the approach turns out to be workable. Thus the proposed method, together with  habitual calculations using NCSM model allows one to calculate simultaneously nearly all properties of ground and excited states of light nuclei. 

The critical point of the approach is a correct representation of the form of the CFF at distances at 
which, first, the nuclear interaction is negligible and, second, the "memory" of the exchange effects 
remains exclusively in the exchange kernel matrix $ ||N_{nn'} ||$.

\section{Technical details of the calculations}

As it was declared above the goal of the current study is a simultaneous description of the energies and the alpha-decay widths of a large list of the positive parity  states of $^8$Be nucleus. It is important to present special features of the application of the proposed general approach  to the stated problem and to justify their need.

First, even large-scale high-precision calculations could not completely reproduce the data of the spectroscopic tables concerning a certain nucleus. On the one hand many unknown levels appear in this calculations. On the other hand some well-known states turn out to be "lost" or their energies are significantly shifted. It depends on the choice of the NN-potential and/or the size of the basis. To make the pattern of the theoretical results more understandable we perform the computations using two well-tested potentials Daejeon16 \cite{dj16} and JISP16 \cite{jisp1}  in the current study. The NCSM calculations of the energies and WFs were carried out with the use of the Bigstick code \cite{bigstick} which is convenient for use on multiprocessor computing clusters.

\end{multicols}

\begin{flushleft}
\begin{table}
\caption{ Total binding energies of lower states of $^8$Be nucleus \newline obtained with the use of the JISP16  potential with $\hbar \omega$ = 22.5 MeV.}\label{jisp1}
\begin{tabular*}{0.7\textwidth}{c|c|lllll}
\hline\hline
 $J^\pi$ & model&N$_{max}^{tot}$ = 4& N$_{max}^{tot}$ = 6 & N$_{max}^{tot}$ = 8 &  N$_{max}^{tot}$ = 10 &  N$_{max}^{tot}$ = 12\\
\hline
0$^+$&NCSM &  22.207 & 34.930 & 44.624 & 49.679 & 52.247 \\
&RGM-like & 21.765 & 23.622 & 29.755 & 33.191 & ---  \\
\hline
2$^+$&NCSM & 18.230 & 30.539 & 40.650 & 45.849 & 48.479 \\
&RGM-like &--- & 13.894 & 21.407 & 26.258 & ---  \\
\hline
4$^+$&NCSM & 	 10.160 & 21.008 & 31.470 & 36.723 & 39.450 \\
&RGM-like & --- & --- & 8.911 & 13.966 & --- \\
\hline\hline
\end{tabular*}
\end{table}
\end{flushleft}

\begin{flushleft}
\begin{table}
\caption{The same as in Table \ref{jisp1} obtained with the use of the Daejeon16  potential with $\hbar \omega$ = 15 MeV.}\label{dae1}
\begin{tabular*}{0.7\textwidth}{c|c|lllll}
\hline\hline
 $J^\pi$ & model&N$_{max}^{tot}$ = 4& N$_{max}^{tot}$ = 6 & N$_{max}^{tot}$ = 8 &  N$_{max}^{tot}$ = 10 &  N$_{max}^{tot}$ = 12\\
\hline
0$^+$&NCSM & 36.204 & 46.467 & 52.169 & 54.618 & 55.721 \\
&RGM-like  & 33.999 & 38.662 & 44.151 & 45.723 & ---  \\
\hline
2$^+$&NCSM & 32.901 & 42.818 & 48.531 & 51.071 & 52.246 \\
&RGM-like & --- & 32.311 & 38.673 & 41.031 & ---  \\
\hline
4$^+$&NCSM & 25.462 & 34.085 & 39.774 & 42.417 & 43.784 \\
&RGM-like  & --- & --- & 27.913 & 31.142 & --- \\
\hline\hline
\end{tabular*}
\end{table}
\end{flushleft}

\begin{multicols}{2} 

Second,   it is well known that 0$_1^+$, 2$_1^+$, and 4$_1^+$ states of $^8$Be nucleus  are strongly clustered. The SFs of the $\alpha+\alpha$ channel in  these states determining the contribution of the channel WFs to the total WFs may be found bellow in Tabs. \ref{dae2}, \ref{jisp2}. Nevertheless in our previous work \cite{our2} it was demonstrated that the relatively small contribution of non-clustered configurations in the WFs of the clustered states is crucially important in calculating the total binding energies of these states. The ground state of the discussed nucleus was considered  as an example. A more detail illustration of this fact is represented in the Tabs.  \ref{jisp1}, \ref{dae1} where total binding energies of 0$_1^+$, 2$_1^+$, and 4$_1^+$ states using NCSM and the cluster basis are shown. For the Daejeon16 NN-potential the contribution of the $\alpha+\alpha$ (RGM-like) components of the basis with realistic WFs of $^4$He clusters to the total binding energy is dominating  only for the case of very short basis. The  basis cut off parameter increase results in the fast growth of the contribution of the components of different nature (non-clustered ones).  For the JISP16 NN-potential the contribution of the non-clustered components of the basis to the total binding energy  is rather large even in the case of the short basis.  The reason is that the Daejeon16 interaction is "softer" in the sense of SRG-transformation. This tendency is confirmed by the results of the paper \cite{kv1} in which a very soft potential with the parameter $\lambda$= 1.4 fm$^{-1}$ was used and the difference in the total binding energy NCSM and the RGM calculations turned out to be not so great. Anyway the problem of the total binding energy computations using the realistic but not effective NN-potentials could not be solved in the framework of a pure cluster model. What about the excitation energies of the clustered states of nuclei the situation looks somewhat better but even in such cases it is hard to say that these results are trustworthy. Obviously the RGM-like bases instead of complete NCSM one are unusable for investigations of non-clustered and slightly-clustered states. That is why we use the full-size NCSM basis for the calculations of all values under study: the binding and the excitation energies, the CFFs, and the SFs.  

Third, various computations performed in the framework of NCSM demonstrate that a great basis is necessary to reach a convergency of the values of the total binding energies and the excitation energies of light nuclei in the case that the Daejeon16 and JISP16 interactions are studied. That is why the  basis cut off parameter N$_{max}^{tot}$ = 14 is exploited in the calculations of the energies and the WFs of the $^8$Be nucleus. The size of Slater determinant basis corresponding to N$_{max}^{tot}$ = 14 is about $2 \cdot 10^8$.  The requirements to the basis necessary for an accurate computation of the cluster SFs and CFFs are studied in the current work. Some results of this study concerning the lowest 0$^+$ state of $^8$Be are illustrated by Table \ref{cff}. The coefficients (amplitudes) $K^{nl=0}_{AA_1A_2}$ of CFF expansion onto oscillator functions are presented in the table. These values indicate that the dominant amplitudes of the CFF mostly converge for the basis cut off parameter $N^{max}_{tot}= 12$ and the choice of the cut off parameter $N_{cl}^{max}$of the bases of the cluster WFs weakly affects these quantities. So it is shown that the channel form factor and  channel SF do not depend noticeable on the accuracy of the subsystem description and one can use amplitudes with a relatively small cluster cut off parameter $N_{cl}^{max}$ and, respectively, large quantum number of relative motion for better description of CFF asymptotic range. The use of the large  basis cut off parameter N$_{max}^{tot}$ = 14 to obtain $^8$Be nucleus WFs is, nevertheless, preferable  in the framework of calculations of the decay widths because the matching procedure of the CFF and the asymptotic WF determining a certain decay width requires a realistic description of the former value in the peripheral region. We illustrate this issue bellow.

\end{multicols}

\begin{flushleft}
\begin{table}
\caption{\label{cff} Amplitudes of the $\alpha$-cluster form factor $K_{AA_1A_2}^{nl=0}$ for the ground state of $^8$Be nucleus in various bases.}
\begin{tabular*}{0.735\textwidth}{c|lllllllll}
\hline\hline
$N_{tot}^{max}$ &$N_{cl}^{max}$ & n=0 &n=2 & n=4 & n=6 &n=8 & n=10 & n=12 & n=14\\
\hline
12    & 0 &  0.0   &  0.0   & 0.785  & -0.285 & 0.195 & -0.095 & 0.040 & ---  \\
      & 2 & -0.113 & -0.258 & 0.819  & -0.284 & 0.210 & ---    & ---   & ---  \\
      & 4 & -0.078 & -0.135 & 0.878  & ---    & ---   & ---    & ---   & ---  \\
\hline
14    & 0 &  0.0   &  0.0   & 0.761  & -0.302 & 0.223 & -0.128 & 0.071 & -0.029  \\
      & 2 & -0.107 & -0.249 & 0.795  & -0.306 & 0.243 & -0.139 & ---   & ---  \\
      & 4 & -0.0063 & -0.237 & 0.805 & -0.322 & ---   & ---    & ---   & ---  \\
\hline\hline
\end{tabular*}
\end{table}
\end{flushleft}

\begin{multicols}{2}

Forth, in the present work we are focused  primarily on calculation of asymptotic properties of  $^8$Be states namely the alpha-decay widths. For sub-barrier processes these values are strongly dependent on the decay energy. As it is demonstrated above the experimental total binding energies of $^8$Be nucleus states are well reproduced nearly for all states (see Tabs. \ref{dae2},\ref{jisp2},\ref{abn}). But the decay energy being differential quantity is evidently reproduced with a lower relative precision. Therefore the question whether the achieved so far accuracy of NCSM computations of the level energy over the decay threshold satisfactory to determine such decay widths. The $\alpha+\alpha$-decay of the ground state of $^8$Be  is a good object for the analysis.  In calculation using the Daejeon16 potential the total binding energy of $^4$He is equal to 28.372 MeV whereas the experimental value of it is equal to 28.296 MeV. This leads to the difference in theoretical and experimental resonance energy values for lowest 0$^+$ state $\Delta E$=335 keV.  Such a difference is typical for high-precision calculations of the total binding energy of lower levels of nuclei from the discussed mass region. However the absolute values of the resonance energy look very different. They are: 92 keV (value from data tables) and 427 keV (value calculated using the Daejeon16 interaction one)  respectively. Let us consider the effect of the substitution of both quantities into the formulas of the decay width. The effect is illustrated by Fig. \ref{fig2} showing  the behaviour of irregular  Coulomb WF $G_0 (\eta,k\rho_{in} )$ at the discussed energies. As it is seen the substitution of the calculated energy value would result in overestimation of the decay width of 0$^+$ state by more than two orders of magnitude compared to the substitution of the proper experimental value. This effect is not so drastic but significant for 2$^+$ state and negligible for higher excited states. A clear illustration of that is Fig. \ref{fig3} showing  the behaviour of regular and irregular  Coulomb WFs $F_4 (\eta,k\rho_{in} )$ and $G_4 (\eta,k\rho_{in} )$ at the calculated and the experimental energies characterizing 4$^+$ state ($\Delta E$ =270 keV). In fact the effect is poorly visible in spite of the fact that the discussed level is placed near the top of the potential barrier. Nevertheless in the current paper the experimental resonance energies are used where possible.  On the other hand the figure Fig. \ref{fig3} shows that for states with such energy the regular solutions could not be neglected.

\begin{center}
\includegraphics[width=1.0\linewidth]{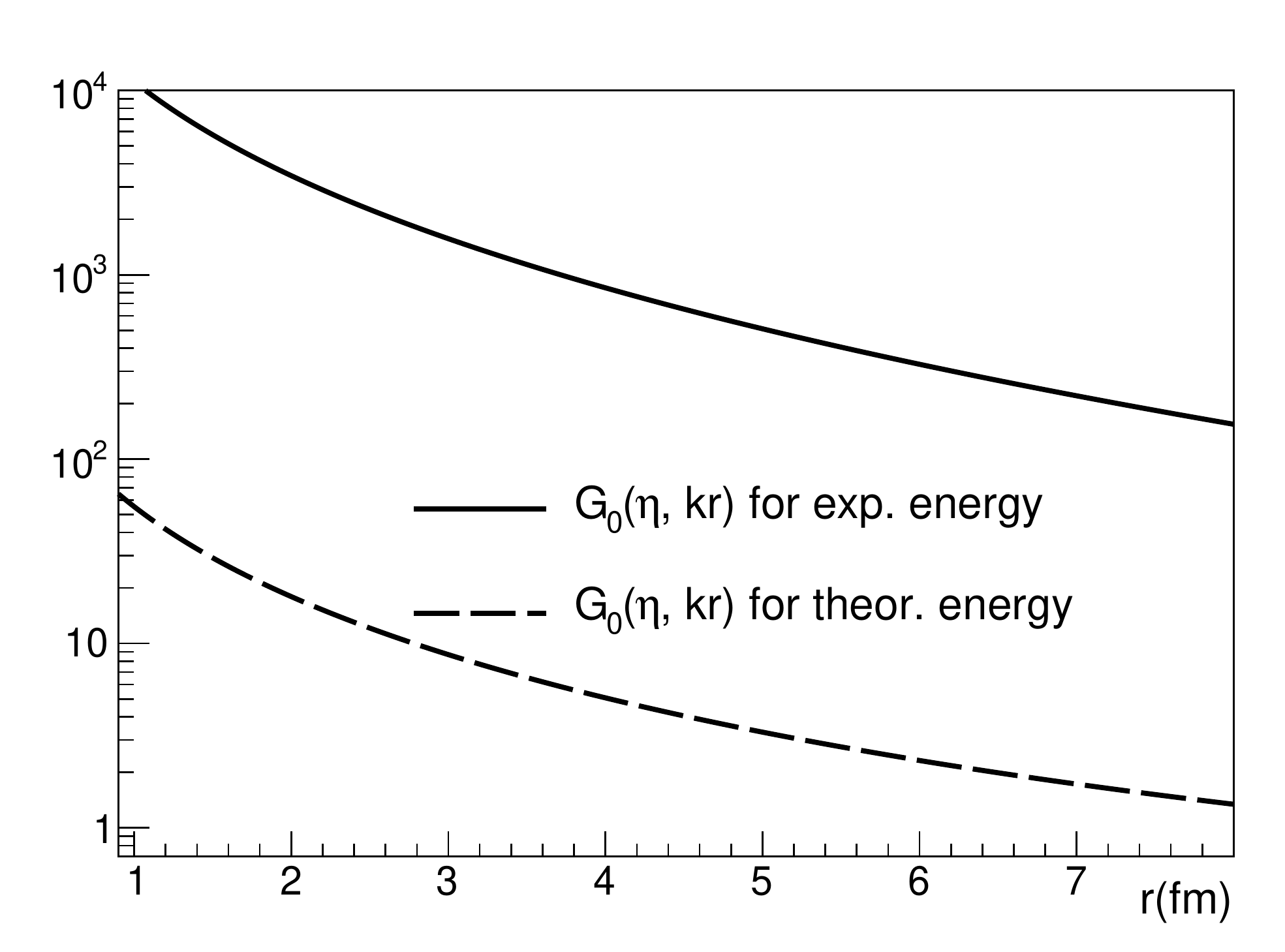}
\figcaption{\label{fig2} The asymptotic irregular WF of $\alpha+\alpha$ channel for the experimental and theoretical decay energies of 0$_1^+$ state of $^8$Be nucleus.}
\end{center}

\begin{center}
\includegraphics[width=1.0\linewidth]{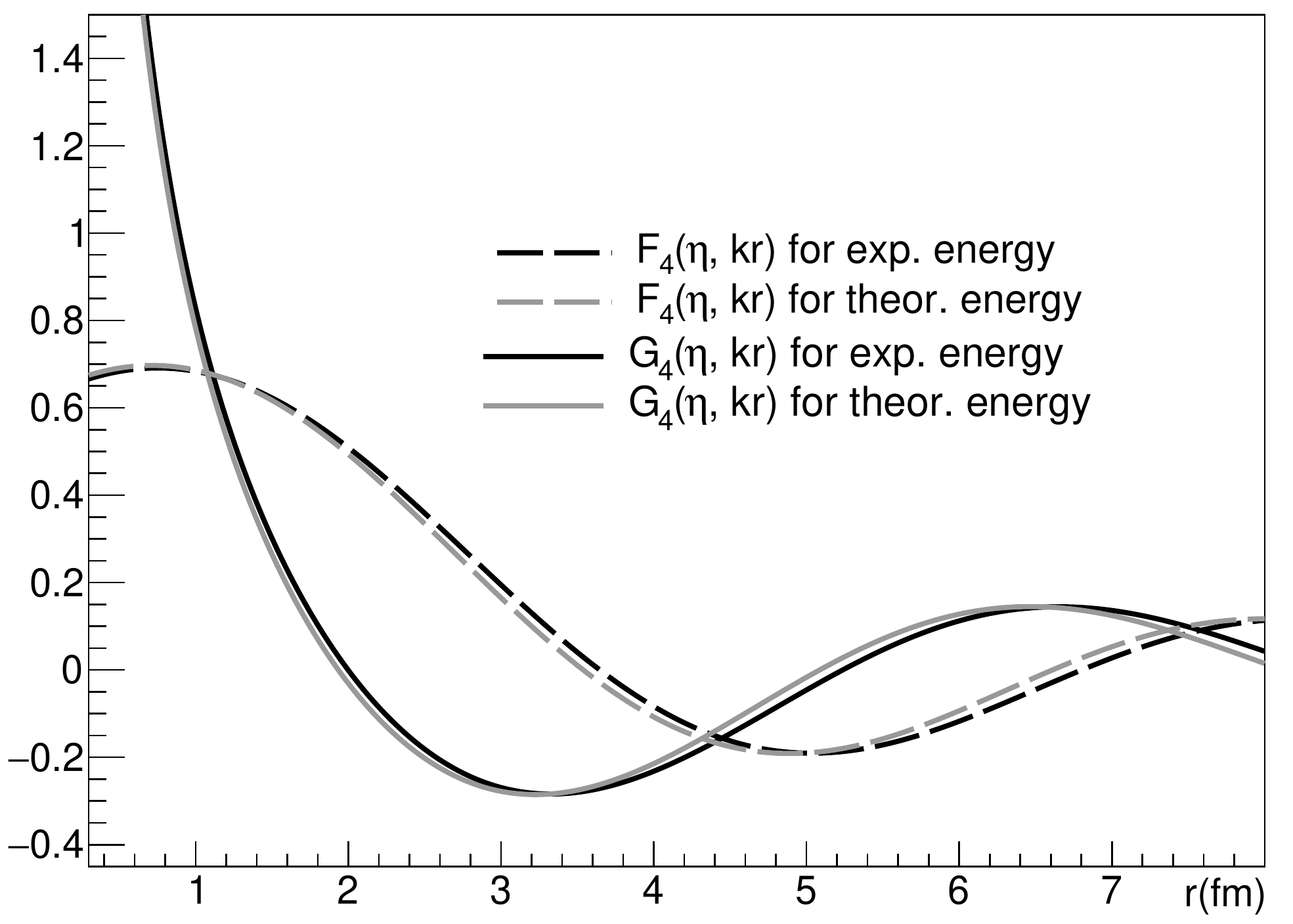}
\figcaption{\label{fig3} The asymptotic regular and irregular WFs of $\alpha+\alpha$ channel for 4$_1^+$ state of $^8$Be nucleus decay.}
\end{center}

It makes sense to present some peculiarities of matching procedure in use. The CFFs as functions of distance between the centres of mass of the $\alpha$ clusters  for the  states 0$_1^+$, 2$_1^+$, and 4$_1^+$ of $^8$Be nucleus which are computed using the Daejeon16 NN-potential are presented in Fig. \ref{fig1}. The outer maxima of all these functions are located in the range 3.0 $\div$ 3.4 fm. To determine the decay widths this functions should be matched with asymptotic solution. For the first two of them the condition (\ref{eq15}) and consequently (\ref{eq17})  are valid. For the third one the width is determined by expression (\ref{eq20}).

\begin{center}
\includegraphics[width=1.0\linewidth]{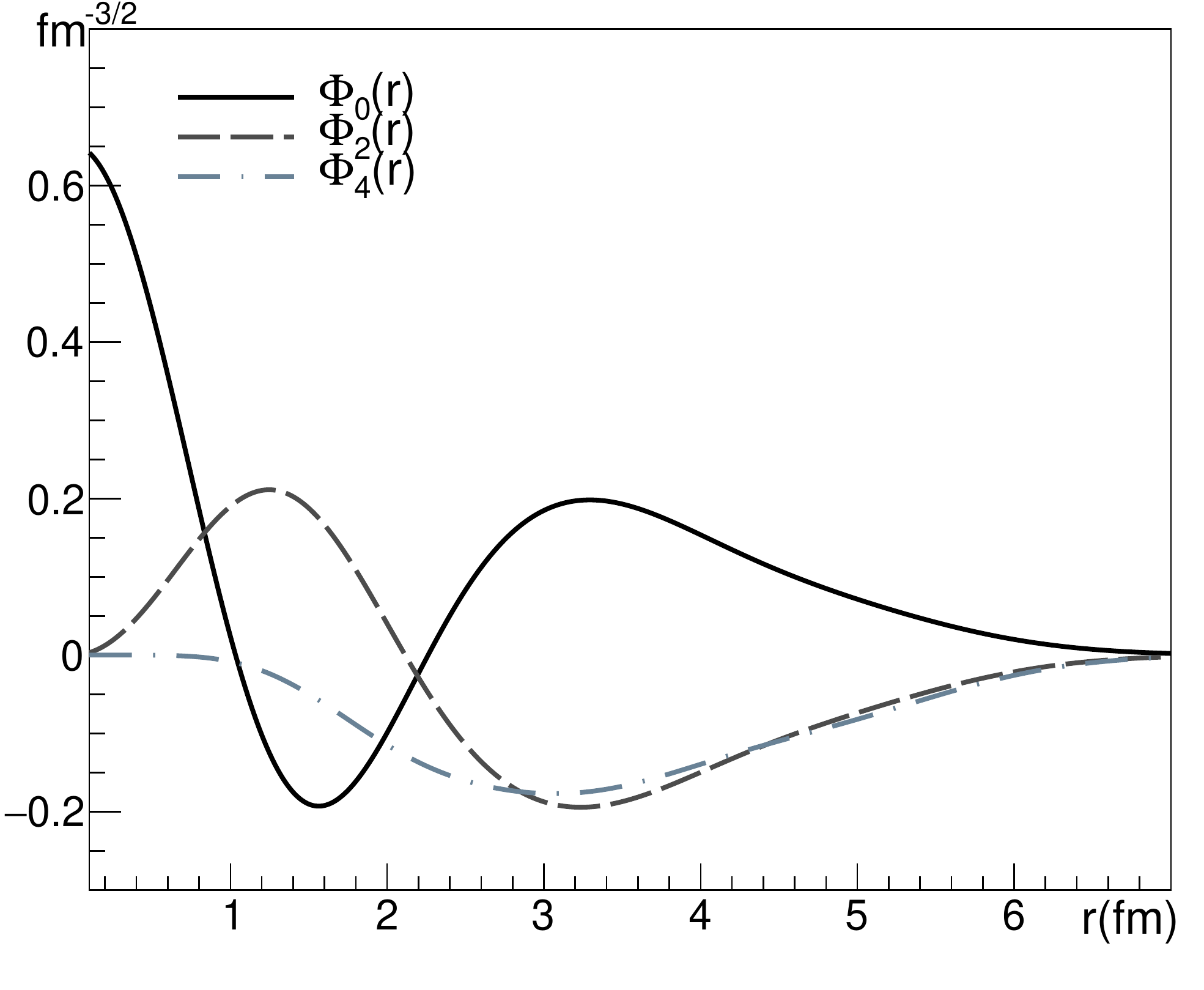}
\figcaption{\label{fig1} CFF for  0$_1^+$, 2$_1^+$, and 4$_1^+$ states of $^8$Be nucleus.}
\end{center}

\begin{center}
\includegraphics[width=1.05\linewidth]{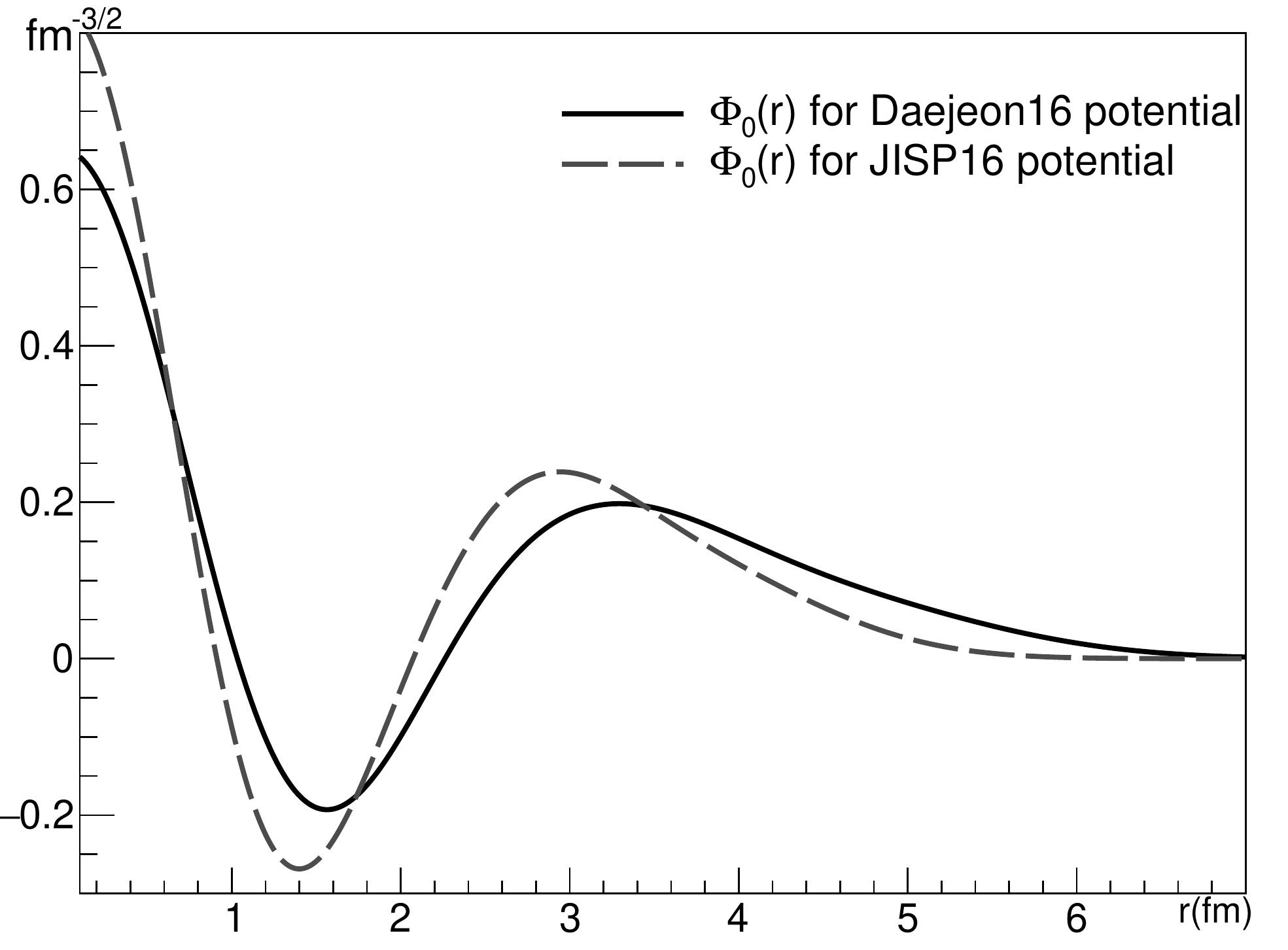}
\figcaption{\label{fig10} CFF for the ground state of $^8$Be nucleus.}
\end{center}

The procedure of calculation of decay channel asymptotic properties was discussed above (see Eqs. (\ref{eq15} -- \ref{eq20})). The matching of asymptotic solution with CFF is performed in the coordinate space, so CFF as a function of relative motion is defined. An example demonstrating difference in the form of the CFFs obtained using the Daejeon16 and JISP16 NN-interactions is presented in Fig. \ref{fig10}. The ground state of  $^8$Be nucleus is considered. The outer maxima of the curves spaced by about 0.4 fm. As we demonstrate bellow such a noticeable difference in the shape of the functions does not necessarily result in a  significant difference of the decay widths.

The reliability of the matching procedure may be tested by the analysis of the behaviour of the ratio of the CFF to the asymptotic WF $ \Phi _A^{c_\kappa  } (\rho_{m})/G_l (\eta,k\rho _{m})$ or corresponding ratio from Exp.  (\ref{eq17}) near the matching point. In Fig.  \ref{fig4} the discussed behavior for the lowest 0$^+$, 2$^+$ and 4$^+$ states is pictured. The matching points characterizing the decay process of these states turn out to be located at the distances 3.92, 3.96, and 4.34 fm respectively. As is obvious from the figure all these ratios vary only slightly near their matching points. The behaviour shown here provides reason enough to conclude that the CFF computed in the framework of the chosen NCSM basis takes the form of the asymptotic WF in the presented range of distances. In other words the required asymptotic  is achieved. Minor variations of the asymptotic WFs with the change of the decay energy $\Delta E$ =270 keV (see Fig.\ref{fig3}) together with the stability of the discussed ratio in the vicinity of the matching point are properly confirm the reliability of the procedure used in the current work to calculate the alpha-decay widths.

\begin{center}
\includegraphics[width=1.05\linewidth]{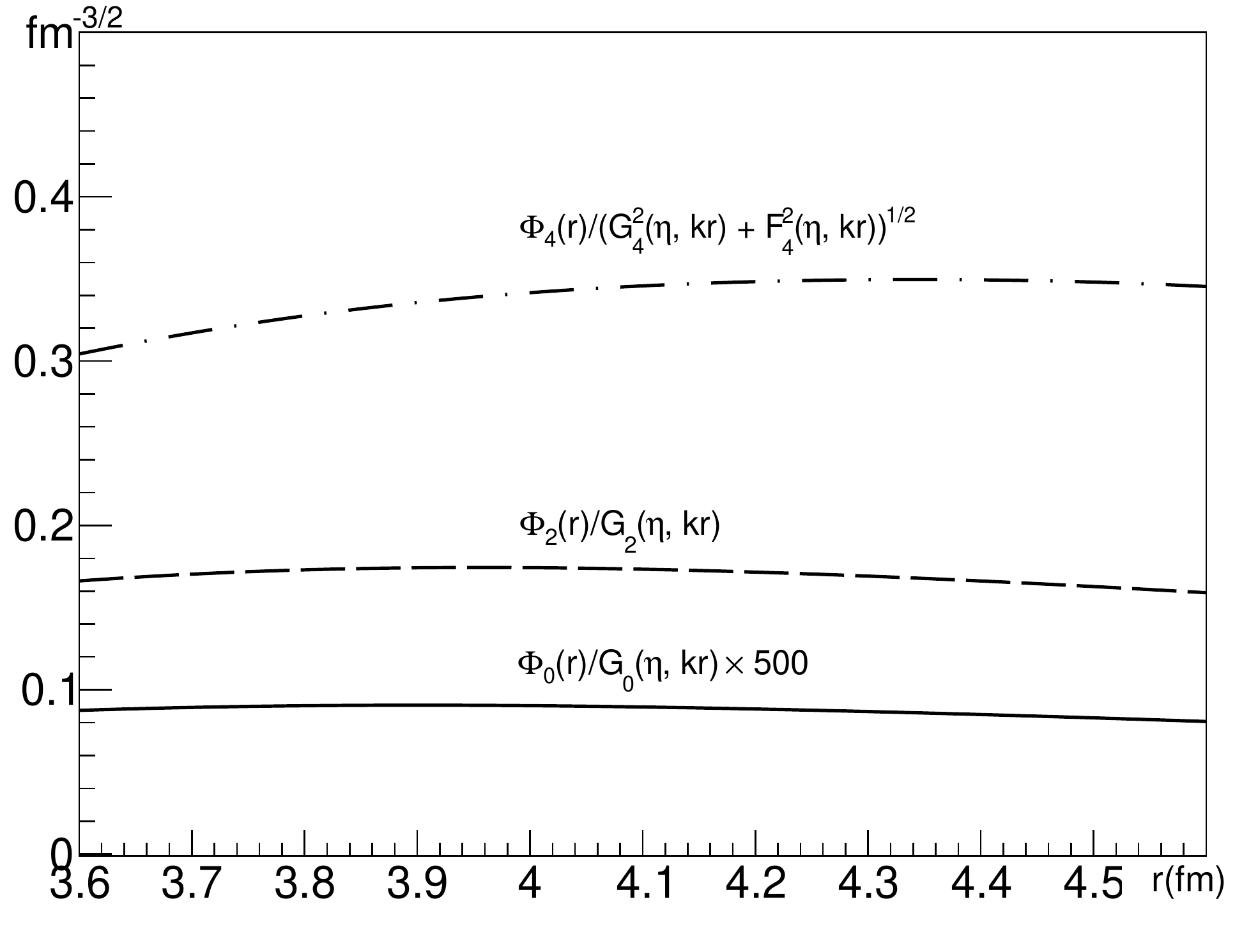}
\figcaption{\label{fig4} The ratio  ($\Phi_{l}(r)/G_l(\eta, kr)$ for   0$_1^+$, 2$_1^+$, and 4$_1^+$ states of $^8$Be nucleus.}
\end{center}

\section{Results and discussion}

In the framework of the just presented scheme the computations of the total binding and excitation energies, statistical weights of the components with the isospin $T=1$, and alpha-decay widths of the positive parity states  of  $^8$Be nucleus with the excitation energy ranging up to about 25 MeV were carried out. Thus it seems reasonable to qualify the current work as an effort to describe almost complete set of the spectroscopic data related to the discussed states theoretically. The results of the computations  are presented in Tables \ref{dae2}, \ref{jisp2}, and \ref{abn}.  The experimental data for all levels whose spin and parity are determined are shown too. 

As the data of Table \ref{dae2} suggest, the calculations based on the Daejeon16 NN-potential reproduce the experimental value of the total binding energy of the ground state of the discussed nucleus $E_b=56.50$ MeV with a high precision. What about the total binding energy of the ground state of  $^8$Be nucleus calculated using the JISP16 NN-potential (see Table  \ref{jisp2}) this value is  2.73 MeV less compared to the just presented experimental one.  As it is known the values of the total binding energies of nuclei from the the mass region A$ \sim 8$ more or less close to the experimental ones may be also achieved in the calculations with the second mentioned potential. However it is necessary for that, first, to exploit a supercomputer and, second, to extrapolate the values obtained for a number of different values of $\hbar\omega$ to higher values of  basis cut off parameter N$_{max}^{tot}$ using a special technique. So the Daejeon16 potential has the advantage of more rapid convergence of the total binding energy. 

Consideration of the excitation energies of $^8$Be nucleus states themselves shows very good agreement of the results obtained, respectively, using the Daejeon16 potential ($E^*_{dae}$) and the JISP16 interaction ($E^*_{jisp}$) with each other as well as those and others with the tabulated experimental data $E^*_{exp} $. For levels of abnormal parity 1$^+$ and 3$^+$ it is evident from Table \ref{abn}.  Just two quantities obtained by the use of the JISP16 potential $E^*_{jisp}$ differ by  more than 1 MeV from the experimental ones. There is no one example of such difference among the results of Daejeon16-based calculations. In general a moderate advantage of the latter approach manifesting in the smaller root-mean square deviation from the experiment is seen. The discussed table shows that excitation energy of the predicted states of abnormal parity for two exploited interactions differ by about 1 MeV or less. Only one state of this parity predicted by JISP16-based computations (3$_3^+$) is not reproduced by Daejeon16-based  ones. 

The excitation energies of normal-parity  states of  $^8$Be nucleus contained in various nuclear data tables are presented in the fifth columns of Tables \ref{dae2}, \ref{jisp2}. For the first glance it is easy to establish one-to-one correspondence between each of these states and any state obtained in the framework of the computations performed using  the Daejeon16 NN-potential or the JISP16 interaction as it is presented in Tables \ref{dae2}, \ref{jisp2}. Moreover both these interaction models result in a good quantitative agreement of the theoretical data and the experimental ones. There are three examples of significant but not critical discrepancy of the energy values for Daejeon16-based investigations: about 1 MeV -- for levels 0$_3^+$ and 2$_5^+$, and more than 2 MeV -- for level 4$_3^+$. A similar pattern characterizes JISP16-based investigations:  difference about 1 MeV is observed for level 4$_1^+$ and more than 2 MeV -- for level 0$_3^+$. As it is the case in consideration of  the abnormal parity states a modest advantage of the Daejeon16-based approach is seen. Some levels predicted in one and the other of the discussed approaches allow one to establish one-to-one correspondence between them. Indeed, a good correlation of the excitation energies of the levels with identical quantum numbers predicted by the Daejeon16-based and JISP16-based approaches respectively takes place for four examples: 2$_6^+ \leftrightarrow 2_5^+$, 2$_8^+ \leftrightarrow 2_7^+$, 0$_5^+ \leftrightarrow 0_4^+$ and 4$_4^+ \leftrightarrow 4_3^+$. 

Naturally the isospin quantum number is a good identifier of nuclear states. Because of that the statistical weights of the components of WFs  with the isospin $T=1$ are also calculated. They are contained in Tables \ref{dae2}, \ref{jisp2}, and \ref{abn}. In the current work we prefer to demonstrate the values of multipliers of these components (they are denoted by symbol $\bar T$) but not their squares. All these values are concentrated near zero or unity therefore they provide an additional and, as it is well-seen from Tables \ref{dae2} and \ref{jisp2}, important means to classify the states in a complicated spectrum and establish the correspondence between calculated and measured levels. The information on weights of the components with certain isospin play an independent role in studies of Coulomb and some exotic effects in nuclei. 

The duplicated prediction of the states of both normal and abnormal parity in two different approaches (see above) together with a good description of a lot of known levels gives ground to propose experiments aimed at search for the predicted states.  This proposal looks promising because detection of levels predicted theoretically (perhaps a part of them) would be a justification of capability of the high-precision theoretical approaches in the spectroscopic studies of light nuclei. That seems to be especially important in the studies of exotic light nuclei. 

A major novelty of the presented work is the computations of the alpha-decay widths of all decaying (i. e. normal parity) states  of  $^8$Be nucleus. The values of alpha-decay widths of various nuclides being measured or calculated are of fundamental importance in low-energy nuclear physics and nuclear spectroscopy in particular. They determine branching ratios of resonance nuclear reactions with alpha-particle in the entrance or exit channels. The quantities directly connected with them, namely SFs and asymptotic normalization coefficients on the alpha-cluster channels play a significant role in the analysis of nuclear fusion and direct nuclear reactions. 

A knowledge of the discussed widths helps to determine quantum numbers of decaying states and thus to build nuclear spectra. In the present section we demonstrate the capabilities of the approach basing on the computation of the alpha-decay widths in the studies of nuclear spectra on the example of $^8$Be nucleus. The experimental data on the total decay widths are contained in the last columns of  Tables \ref{dae2} and \ref{jisp2}.  The thresholds of proton and neutron decay of $^8$Be nucleus are located at the energies $S_p=17.26 $ and $18.90$ MeV respectively. Thus the alpha-decay width of a lower level coincides with the total width. For higher levels a total widths presented in the spectroscopic tables may serve as upper limits of the corresponding alpha-decay widths. 

The question arises of whether the accuracy of the decay widths computations enough to consider a calculated width as a reliable and therefore able to serve as an identifier of a state characteristics. The tables under discussion demonstrate that sometimes discrepancy of the values extracted from experiments and the ones obtained theoretically turns out to be several times. This discrepancy appears probably due to, first, application of the potentials which are not specially adopted to the calculations of the CFFs and, second, the necessity to use the simple version of the R-matrix theory for highly excited states. To answer the question it is reasonable to take a second look at the range of variation of the widths presented in Tables \ref{dae2} and \ref{jisp2}, both experimental and theoretical. For the calculated values this range is more than 500 times if even the cases of the decay of levels with the isospin $\bar T \sim 1$ and the special case of the ground state are excluded. The same range is a characteristic of the number of the alpha-decay widths extracted from the experiments. Thus occasional coincidences of the values of the decay widths obtained in the calculations and in the measurements are unlikely. This fact gives reliable grounds to use a procedure of comparison of the  theoretically obtained and experimentally extracted decay widths for search for a correspondence between certain states.  It is interesting to note that sometimes even the SF (these values are presented in the fifth columns of Tables \ref{dae2} and \ref{jisp2}) turns out to be satisfactory identifier of a state because the values of SFs and the decay widths are rather strongly correlated.   

This comparison offers complementary possibilities to analyse various nuclear spectra and the  $^8$Be nucleus spectrum in particular. The analysis of the data obtained in the Daejeon16-based calculations leads to the following conclusions. The great decay width of state 4$_3^+$ (5.13 MeV) which  is much more than the total decay width of the known 19.96 MeV state in addition to rather large discrepancy in the excitation energy is evidence that the identification of this state is most likely incorrect. So 4$_3^+$ state presented in the spectroscopic tables is not reproduced in the Daejeon16-based calculations in contrast to the JISP16-based ones. A modest disadvantage of the discussed potential manifest itself in overestimation of the alpha decay width of 2$_4^+$ level and underestimation of the width of 2$_3^+$ state. The decay-width test confirms a good reproduction of the properties of all other known levels. The same analysis of the results of the JISP16-based calculations sheds light on the following problems. Two states: 2$_4^+$ and 2$_6^+$ have great calculated widths. As it is the case for the just discussed 4$_2^+$ level obtained in the Daejeon16-based calculations one could not exclude that these two  levels exist in reality and have not been detected yet because their great widths. At the same time the great widths of these levels show that the levels known from the experiments  which were identified with the discussed ones due to the energy marker turn out to be non-reproduced.  These circumstances do not violate the a good general description of $^8$Be nucleus spectrum.

\end{multicols}

\begin{table}
\caption{The spectrum of $^8$Be nucleus calculated using the Daejeon16 potential with $\hbar\omega$= 15.0 MeV \newline
and the experimental data from \cite{exp8be}, $^*$ -- differing data form \cite{toi}.}\label{dae2}
\begin{tabular}{ccccccc}
\hline\hline\noalign{\smallskip} $J^\pi,\bar T$ &$ E_{bind}$ MeV &$E^*_{dae}$ MeV&$E^*_{exp} $MeV($T$)  &SF&$\Gamma_{th}$ MeV&$\Gamma_{exp}$ MeV  \\
\hline\noalign{\smallskip}
0$_1^+,0$&56.25&0.0 &0.0 (0)&0.879&7.29 eV&5.57 (6.8)* eV \\
2$_1^+,0$&52.85&3.40 &3.03$\pm$0.01  (0)&0.849&1.17 &1.513 \\
4$_1^+,0$&44.63&11.62 &11.35$\pm$0.15  (0)&0.792&2.41 &3.5 \\
0$_2^+,0$&44.54 &11.71 & ---&0.813&8.86&---\\
2$_2^+,0.001$&42.09&14.16 & ---&0.715&3.57 &---\\
2$_3^+,0.971$&39.65&16.59 &16.626$\pm$0.003  (0+1)&0.0025&0.019&0.108 \\
2$_4^+,0.078$&39.05&17.19 &16.922$\pm$0.003  (0+1)&0.354&0.416&0.074 \\
4$_2^+,0.001$&37.48&18.76 & ---&0.288&3.39 &---\\
2$_5^+,0.065$&35.02&21.22 &20.1$\pm$0.01 (0) &0.0459&0.434 &0.8 (1.1)\\
0$_3^+,0.852$&35.01&21.23 &20.2$\pm$0.01  (0)&0.0208&0.056&0.7 ($\le$1)\\
0$_4^+,0.315$&34.44&21.80 & ---&0.0610&0.092&---\\
2$_6^+,0.966$&34.27&21.97 & ---&0.0039&0.002&---\\
4$_3^+,0.007$&34.24&22.00 &19.86$\pm$0.05  (0)&0.441&5.13&0.7\\
2$_7^+,0.028$&33.57&22.70 &22.2 (0)&0.059&0.135&0.8 \\
2$_8^+,0.996$&33.22&23.02 & ---&0.001&0.004 & ---\\
0$_5^+,0.017$&32.91&23.33 & ---&0.215&1.71 & ---\\
4$_4^+,0$.997&32.69&23.55 & ---&0.0009&0.009& ---\\

\noalign{\smallskip}\hline\hline
\end{tabular}
\end{table}

\begin{table}
\caption{The same data as in Table \ref{dae2} for the JISP16 potential with $\hbar\omega$= 22.5 MeV.}\label{jisp2}
\begin{tabular}{cccccccc}
\hline\hline\noalign{\smallskip} $J^\pi,\bar T$ &$ E_{bind}$ MeV &$E^*_{jisp}$ MeV&$E^*_{exp} $MeV($T$)  &SF&$\Gamma_{th}$ MeV&$\Gamma_{exp}$ MeV  \\
\hline\noalign{\smallskip}
0$_1^+,0$&53.77&0.0 &0.0 (0)&0.841&6.72 eV&5.57 (6.8)* eV \\
2$_1^+,0$&50.11&3.66 &3.03$\pm$0.01  (0)&0.803&1.08 &1.513 \\
4$_1^+,0$&41.28&12.50&11.35$\pm$0.15  (0)&0.729&1.65 &3.5 \\
2$_2^+,0.987$&37.11&16.66 &16.626$\pm$0.003  (0+1)&0.0003&0.005&0.108 \\
2$_3^+,0.038$&36.45&17.33 &16.922$\pm$0.003  (0+1)&0.018&0.305&0.074 \\
0$_2^+,0.002$&34.80 &18.98 & ---&0.698&10.45& ---\\
4$_2^+,0.002$&33.87&19.91 &19.86$\pm$0.05  (0)&0.022&0.249&0.7\\
2$_4^+,0.008$&33.06&20.72 &20.1$\pm$0.01 (0) &0.166&2.91 &0.8 (1.1)*\\
2$_5^+,0.995$&31.87&21.90 & ---&0.0045&0.079& ---\\
0$_3^+,0.986$&31.55&22.23 &20.2$\pm$0.01  (0)&0.0086&0.110&0.7 ($\le $1)*\\
2$_6^+,0.005$&31.45&22.33 &22.2 (0)&0.354&5.53&0.8\\
2$_7^+,0.999$&30.16&23.62& ---&0.0008&0.0011& ---\\
0$_4^+,0.039$&30.17&23.64 & ---&0.317&2.48& ---\\
4$_3^+,0.999$&29.42&24.36& ---&0.00015&0.0026& ---\\
2$_8^+,0.004$&28.72&25.06 &---&0.270&3.67 & ---\\
4$_4^+,0.001$&27.58&26,20& ---&0.216&3.74& ---\\

\noalign{\smallskip}\hline\hline
\end{tabular}
\end{table}

\begin{multicols}{2}

Let us come back to the analysis of the predicted levels of normal parity. The triple of 0$_2^+$, 2$_2^+$,  4$_2^+$ states with great alpha-decay widths contained in Table \ref{dae2} attracts the major attention. It looks like a typical rotational band with the rotational quantum $\hbar^2/2\mu r^2 \sim 0.38$ MeV. The SFs and the alpha-decay widths of these states confirm that they are strongly clustered.  Surprisingly a similar rotational band may be found in the spectrum obtained in the JISP16-based computations. It is the triple 0$_2^+$, 2$_4^+$,  4$_4^+$ characterizing by the rotational quantum $\sim 0.33$ MeV. At the same time a great difference in the location of these bands in  $^8$Be nucleus spectrum takes place. It is important to note that this band could not be found in the absence of the data on the alpha-decay widths. This duplicated prediction makes the situation intriguing. In our opinion it would be interesting to detect such a new band of alpha-clustered states and to determine real excitation energy of its members.

There are a number of other states possessing noticeable cluster properties as predicted in the both interaction models (0$_5^+ \leftrightarrow 0_4^+$) as found in one of the two approaches (4$_3^+$ -- for the Daejeon16-bases studies), (2$_4^+$, 2$_6^+$, 2$_8^+$ -- for JISP16-based ones). Experimental studies of the $^8$Be nucleus spectrum in the excitation energy area in which all mentioned states are located, and, in principle, the spectrum as a whole, are of interest not only as a contribution to the spectroscopic information array and physics of nuclear clustering. These investigations may serve as a test to check the quality of various NN-potentials exploiting in ab initio calculations.

Perhaps a popular experimental approach aimed to measure the cross-sections of elastic scattering of alpha-particles from various light nuclei --  so-called  Thick Target Inverse Kinematics technique proposed in Refs. \cite{art}, \cite{goldb} (a detail description can be found in Ref. \cite{avil}) --  adopted  for the discussed purposes would be  convenient for the proposed measurements.

In conclusion let us list the basic points of the performed investigations.

\noindent
1. A method allowing one to carry out the simultaneous ab initio calculations of the total binding and excitation energies, statistical weights of the components which are characterized by certain value of the isospin together with  the  quantities which determine the degree of the alpha-clustering: SFs, CFFs as well and alpha-decay widths is developed. 

\noindent
2. The ab initio theoretical studies of the properties of $^8$Be nucleus states located in a wide range of the excitation energies, both clustered and non-clustered, are carried out for the first time.

\begin{flushleft}
\tabcaption{The values of the isospin and the excitation energy for  the states of abnormal  parity for the Daejeon16 and JISP16 potentials.}\label{abn}
\footnotesize
\begin{tabular*}{0.47\textwidth}{cccccc}
\toprule $J^\pi$ &$\bar T_{dae}$ &$\bar T_{jisp}$& \begin{tabular}{c} $E^*_{dae}$\\ MeV\\ \end{tabular} & \begin{tabular}{c} $E^*_{jisp}$\\ MeV\\ \end{tabular} & \begin{tabular}{c} $E^*_{exp}$\\ MeV($T$)\\ \end{tabular} \\
\hline

1$_1^+$&0.994&0.996&18.01 &18.43& \begin{tabular}{c} 17.640$\pm$ \\ 0.010  (1) \\ \end{tabular}\\
3$_1^+$&0.993&0.998&18.89&19.36 & \begin{tabular}{c} 19.07$\pm$ \\0.03  (1) \\ \end{tabular}\\
1$_2^+$&0.036&0.020&19.14&19.72 & \begin{tabular}{c} 18.150$\pm$ \\0.004  (0) \\ \end{tabular}\\
3$_2^+$&0.023&0.007&19.72&20.46 & \begin{tabular}{c} 19.235$\pm$ \\0.010  (0) \\ \end{tabular}\\
1$_3^+$&0.992&0.997&21.13&22.46 &---\\
1$_4^+$&0.020&0.008&21.33 &21.64&---\\
3$_3^+$&---&0.007&---&23.82&---\\
1$_5^+$&0.994&0.997&23.31&24.36&24.038 \\
\bottomrule
\end{tabular*}
\end{flushleft}

\noindent
3. For the of lowest  rotational band 0$^+_1$, 2$^+_1$ and 4$^+_1$ of $^8$Be nucleus it is demonstrated that  non-clustered components of the WFs of these states make a great contribution to the total binding energy in spite of a small statistical weight of these components in each of the WFs. 

\noindent
4. In the majority of instances the results of the computations of the listed characteristics turn out to be in a good agreement with the tabulated spectroscopic data. 

\noindent
5. It is shown that the alpha-decay width is a good characteristic to identify various nuclear states and to establish a correspondence between observed and calculated nuclear levels.

\noindent
6. A number of levels, both manifesting strongly  and not showing noticeable alpha-clustering properties, which are not found experimentally so far are predicted. The most interesting is two-way prediction of the second rotational band  of $^8$Be nucleus which is strongly clustered.  Corresponding verification experiment is proposed. 

Finally good prospects of ab initio approaches in the studies of a lot of characteristics of light nuclei spectra, interpretation of the properties of known nuclear states as well as prediction of levels which are not observed to date and their characteristics are shown.

\vspace{10mm}

\acknowledgments{
Authors are grateful to A. M. Shirokov , A. I. Mazur and I. A. Mazur for the realistic JISP16 and Daejeon16 NN-potentials matrixes provided by them and to Calwin W. Johnson for his high-performance shell model code Bigstick and assistance in NCSM calculations.}

\end{multicols}

\vspace{10mm}

\begin{multicols}{2}

\end{multicols}

\clearpage
\end{CJK*}
\end{document}